\documentclass[letterpaper, 10 pt, conference]{ieeeconf}  

\IEEEoverridecommandlockouts                              
\usepackage{ntheorem}
\usepackage[T1]{fontenc}
\usepackage{mathtools}
\usepackage{fixltx2e}

\makeatletter
\renewcommand*\env@matrix[1][*\c@MaxMatrixCols c]{%
  \hskip -\arraycolsep
  \let\@ifnextchar\new@ifnextchar
  \array{#1}}
\makeatother
\usepackage{graphicx}      
\usepackage{newlfont}
\usepackage{dsfont}
\usepackage{amssymb,amsmath,amsfonts}
\usepackage{epstopdf}
\usepackage{empheq}
\usepackage[dvipsnames]{xcolor}
\usepackage{cuted}
\usepackage{slashbox}

\newcommand{\Real}{{\mathds R}} 
\newcommand{\Nat}{{\mathds N}} 
\newtheorem{definition}{Definition}{}
\newtheorem{corollary}{Corollary}{}
{}
\newtheorem{theorem}{Theorem}{}
\newtheorem{remark}{Remark}{}
\newtheorem{lemma}{Lemma}{}

\newtheorem{assumption}{Assumption}{}
\newtheorem{problem}{Problem}{}

\makeatletter
\g@addto@macro\normalsize{%
  \setlength\abovedisplayskip{1pt}
  \setlength\belowdisplayskip{1pt}
}
\makeatother

\title{\LARGE \bf
Privacy Against Adversarial Classification in Cyber-Physical Systems
}

\author{Carlos Murguia and Paulo Tabuada
\thanks{Carlos Murguia is with the Department of Mechanical Engineering, Eindhoven University of Technology, The Netherlands; and Paulo Tabuada is with the Department of Electrical Engineering, University of California, Los Angeles, USA. Emails: c.g.murguia@tue.nl, \& tabuada@ee.ucla.edu.}
\thanks{This work was partially supported by the NSF awards 1740047 and 1705135, and by the UC-NL grant LFR-18-548554}
}

\begin{document}

\maketitle
\thispagestyle{empty}
\pagestyle{empty}

\begin{abstract}
For a class of Cyber-Physical Systems (CPSs), we address the problem of performing computations over the cloud without revealing private information about the structure and operation of the system. We model CPSs as a collection of input-output dynamical systems (the system operation modes). Depending on the \emph{mode} the system is operating on, the output trajectory is generated by one of these systems in response to driving inputs. Output measurements and driving inputs are sent to the cloud for processing purposes. We capture this ``processing'' through some function (of the input-output trajectory) that we require the cloud to compute accurately -- referred here as the \emph{trajectory utility}. However, for privacy reasons, we would like to keep the mode private, i.e., we do not want the cloud to correctly identify what mode of the CPS produced a given trajectory. To this end, we \emph{distort} trajectories before transmission and send the corrupted data to the cloud. We provide mathematical tools (based on output-regulation techniques) to properly design distorting mechanisms so that: 1) the original and distorted trajectories lead to the \emph{same utility}; and the distorted data leads the cloud to \emph{misclassify the mode}.
\end{abstract}


\section{Introduction}

In a hyperconnected world, scientific and technological advances have led to an overwhelming amount of user data being collected and processed by hundreds of companies over the cloud. Companies mine and classify this data to provide personalized services and advertising. However, these new technologies have also led to an alarming widespread loss of privacy in society. Depending on the adversaries’ resources, they may infer critical (private) information about the operation of systems from public data available on the internet and unsecured/public servers and communication networks. A motivating example of this privacy loss is the data collection, classification, and sharing by the Internet-of-Things (IoT) \cite{WEBER201023}, which is, most of the time, done without the user’s informed consent. Another example of privacy loss is the potential use of data from smart electrical meters by criminals, advertising agencies, and governments, for monitoring the presence and activities of occupants, \cite{Poor1}-\cite{Poor2}. These privacy concerns show that there is an acute need for privacy preserving mechanisms capable of handling the new privacy challenges induced by a hyperconnected world. That is why researchers from different fields (e.g., computer science, information theory, and control theory) have been attracted to the broad research area of privacy and security of Cyber-Physical Systems (CPSs), see, e.g., \cite{Farokhi1}-\nocite{JRuths1}\nocite{JRuths4}\nocite{Ahmed2017}\nocite{Farokhi2}\nocite{FAROKHI3}\nocite{Murguia2017d}\nocite{Pasqualetti_1}\nocite{Carlos_Justin2}\nocite{Murguia2017d}\nocite{Ozarow}\nocite{Carlos_Justin1}\nocite{Fawaz}\nocite{Hashemil2017}\nocite{Carlos_Justin3}\nocite{Sahand2017}\nocite{Carlos_Iman1}\nocite{Tianci2}\nocite{Jerome1}\nocite{Conf_with_Poor}\nocite{FAROKHI2019275}\nocite{Mitra1}\nocite{Takashi1}\nocite{HaleM}\nocite{Murguia2020}\cite{Alim}.\\[.5mm]

In this manuscript, for a class of Cyber-Physical Systems (CPSs), we address the problem of performing computations over the cloud without revealing private information about the structure and operation of the system. That is, the objective is to have the cloud provide a service by processing system data while preventing it from learning private information. The setting that we consider is the following. The underlying physical part of the system (the system dynamics) consists of a finite collection of $N$ \emph{input-output dynamical systems}, $\Sigma_i$, $i \in \{1,\ldots,N\}$. Depending on the \emph{mode} the system is operating on, sensor measurements are generated by one of these dynamical systems in response to driving inputs. Each subsystem characterizes an operation mode of the CPS. For instance, the operation of fitness trackers is based on different modes (i.e., different dynamical systems) indicating our activity level, e.g., depending whether we are walking, running, or resting, sensors/actuators embedded in the device would provide different data and this data would be consistent with the corresponding dynamical system. That is, we have a dynamical system explaining the data for walking, one for running, and one for resting. Under normal operating conditions, sensor measurements and driving inputs are sent to the cloud for monitoring or processing purposes. However, for privacy reasons, we would like to keep the mode private. To accomplish this, we use knowledge of the system dynamics to appropriately modify sensor measurements and driving inputs generated by/for system $\Sigma_i$ so that the distorted data appears to have been generated by a different \emph{target system}, $\Sigma_j$, $j \neq i$, within the operation modes of the CPS, and we send the distorted data to the cloud. The idea is that if the target system is sufficiently different (in some appropriate sense) from the mode that generated the data, the cloud would incorrectly classify the mode.

Note, however, that we do not want to overly distort the data. The main reason for sharing system data is to have the cloud provide a service by processing it. Usually, there is some function of the sensor data that we would like the cloud to compute accurately--referred here as \emph{the utility function}. For instance, in intelligent transportation systems, we might want the cloud to accurately compute the current highway capacity (the road congestion level) or the shortest route to a destination using, e.g., the average speed of our vehicle (and all other vehicles in the highway). The utility function imposes a constraint on the class of systems that we can use as target systems. Concretely, we aim at modifying input-output data so that: (1) the utility function evaluated at the distorted data equals its value on the original data; and (2) the output trajectory seems to have been generated by the target system in response to driving inputs, i.e., the provided input-output data is consistent with the target system dynamics. We remark that we do not make any assumption on the classification algorithm employed by the cloud. It is unrealistic to assume we know how data is being classified. Hundreds of companies and governmental agencies collect, mine, and classify data from the cloud using advanced machine learning algorithms and data from thousands of users. However, if one is only concerned about misclassifying specific modes of the system dynamics, arguably, mapping output trajectories and driving inputs to a different mode would lead to incorrect classification if the model of the modes that we use to design the distorting mechanism is accurate enough to capture the true dynamic behavior of the system.

Most of the work related to privacy of dynamical systems deals with keeping the system state private (in some appropriate sense) when output measurements and the system model are disclosed for processing purposes, see, e.g., \cite{Jerome1}-\nocite{Conf_with_Poor}\nocite{FAROKHI2019275}\nocite{Mitra1}\nocite{Takashi1}\nocite{HaleM}\cite{Murguia2020}. All these manuscripts follow stochastic formulations where the objective is twofold: 1) To quantify the potential information leakage given a privacy metric (e.g., based on differential privacy \cite{Dwork} or information-theoretic \cite{Cover}); and 2) To design randomizing mechanisms to distort data so that the distorted disclosed data provides prescribed privacy guarantees. In this manuscript, we address a fundamentally different problem. First, we consider fully deterministic systems and thus stochastic privacy metrics do not make sense in our setting. Secondly, we are not concerned with privacy of the system state per se, but it is the mode the system is operating on what we want to keep private. Because we consider LTI dynamics for each mode, all the input-output data that a mode can generate forms a linear subspace (referred here as the mode \emph{behaviour}). So, instead of looking for the probability distribution of the noise to inject (as it is usually done in stochastic formulations), we seek distorting mechanisms, based on system-theoretic tools (output regulation), that maps data from the actual mode behaviour into the behaviour of a different \emph{target mode}, while maintaining its utility invariant.

There are results dealing with privacy of deterministic dynamical systems in the literature -- mostly using encryption/coding techniques \cite{Alim},\cite{Carlos_Farhad_1}\nocite{FAROKHI201713}\nocite{Fujita}\nocite{KIM2016175}\nocite{DARUP2018535}\nocite{Yankai}-\cite{Alexandru}. These techniques rely on two objects: the encryption scheme itself; and an algorithm that can be used to perform the required computations \emph{over the encrypted data}. Both, the encrypted data and the algorithm, are shared with the cloud \emph{but not the} \emph{decryption key}. The cloud then performs computations without having to decrypt, returns the encrypted result, and the user extracts the result using the decryption key. Results in this manuscript are aligned with these ideas. The difference is that our tools do not rely in computationally expensive encryption techniques -- the proposed scheme is linear and easy to implement. Moreover, our scheme does not need to provide an algorithm to perform computations over the distorted data. The idea is that both the original and distorted data return the same utility. Meaning that the service we require the cloud to provide is invariant under the proposed distorting scheme. To the best of the authors knowledge, the problem addressed here has not been considered before as it is posed here.

\textbf{Notation:} The notation $\text{col}(x_1,\ldots,x_n)$ stands for the column vector composed of the elements $x_1,\ldots,x_n$. This notation is also used in case the components $x_i$ are vectors. The $n \times n$ identity matrix is denoted by $I_n$ or simply $I$ if $n$ is clear from the context. Similarly, $n \times m$ matrices composed of only ones and only zeros are denoted by $\mathbf{1}_{n \times m}$ and $\mathbf{0}_{n \times m}$, respectively, or simply $\mathbf{1}$ and $\mathbf{0}$ when their dimensions are clear. Finite sequences of vectors are written as $x^N \coloneqq (x(1)^{\top},\ldots,x(N)^{\top})^{\top} \in \Real^{Nn}$ with $x(i) \in \Real^{n}$, and $n,N \in \Nat$. We denote powers of matrices as $(A)^{K} = A \cdots A$ ($K$ times) for $K > 0$, $(A)^{0} = I$, and $(A)^{K} = \mathbf{0}$ for $K < 0$. Matrix $Q^+ \in \Real^{m \times n}$ denotes the Moore--Penrose inverse of $Q \in \Real^{n \times m}$.

\section{System Description and Problem Formulation}\label{description}

We consider a class of cyber-physical systems whose physical part can be modeled by switching discrete-time linear systems of the form:
\begin{subequations}\label{1}
\begin{align}
&\Sigma_ \rho \coloneqq \left\{ \begin{array}{l}\label{1a}
x_\rho(k+1) = A_\rho x_\rho(k) + B_\rho u(k),\\[0.5mm]
\hspace{6.3mm}y_\rho(k) = C_\rho x_\rho(k), \\[0.5mm] \hspace{4.5mm}\rho \in \mathcal{N} \coloneqq \{ 1,2,\ldots,N \},
\end{array} \right. \\
&\hspace{22.25mm}y(k) = y_\rho(k),\label{1b}
\end{align}
\end{subequations}
with time index $k \in \Nat$, state $x_\rho \in \Real^{n_\rho}$, $n_\rho \in \Nat$, output $y \in \Real^{m}$, $m \in \Nat$, input $u \in \Real^{l}$, $l \in \Nat$, and matrices $A_\rho \in \Real^{n_\rho \times n_\rho}$, $B_\rho \in \Real^{n_\rho \times l}$, and $C_\rho \in \Real^{m \times n_\rho}$. It is assumed that, for all $\rho \in \mathcal{N}$, $A_\rho$, $C_\rho$, and $B_\rho$ are known, $(A_\rho,C_\rho)$ is observable, $(A_\rho,B_\rho)$ is controllable, $\text{Im}[C_\rho] = \Real^m$, and $\text{Ker}[B_\rho] = \{ \mathbf{0} \}$. Depending on the \emph{operation mode} of the system, output data is generated by one of the $N$ subsystems in \eqref{1}, i.e., the output of the system at time $k$, $y(k) \in \Real^{m}$, is given by $y(k) = y_\rho(k)$ if the $\rho$-th mode (system $\Sigma_\rho$) is active, $\rho \in \mathcal{N}$. Although $y(k)$ might switch among different modes, we assume (for trajectory classification to actually make sense) that during a window of observations, $k \in \mathcal{K} \coloneqq \{1,2,\ldots,K\}$, $K \in \Nat$, the trajectory $Y^K = \text{col}[y(1),y(2),\ldots,y(K)] \in \Real^{Km}$ is generated  by a single mode, i.e., $Y^K = \text{col}[y_\rho(1),y_\rho(2),\ldots,y_\rho(K)]$, for some $\rho \in \mathcal{N}$, in response to some driving sequence $U^{K-1} = \text{col}[u(1),u(2),\ldots,u(K-1)] \in \Real^{(K-1)l}$. With slight abuse of notation, we often write $Y^K$ as $Y^K_\rho$ to remark that the trajectory has been generated by subsystem $\Sigma_\rho$.

Each operation mode $\rho \in \mathcal{N}$ characterizes a \emph{behaviour} of the system. For instance, in smart devices, we may have modes indicating our activity level. Depending whether we are walking, running, or idle, sensors embedded in the device provide different output trajectories $Y^K_\rho$. Each trajectory would be \emph{consistent} with the dynamical system $\Sigma_\rho$ that produced it. That is, we have a dynamical system $\Sigma_\rho$ explaining the data for walking, one for running, and one for idle. Thus, when we say that an input-output trajectory $(U^{K-1},Y^K)$ is being classified into a mode $\rho \in \mathcal{N}$, we refer to identifying which system $\Sigma_\rho$ in \eqref{1} produced it. To classify the mode, we characterize the set of all input-output trajectories that $\Sigma_\rho$ could produce, over all possible initial conditions $x_\rho(1) \in \Real^\rho$, and then we test if $(U^{K-1},Y^K)$ belongs to this set -- if so, we say that $(U^{K-1},Y^K)$ is classified into mode $\rho$. We refer to this set of input-output trajectories as the \emph{behaviour} of mode $\rho$.

\begin{definition}[Behaviour]
The behaviour $\mathcal{B}_{\rho} \subseteq \Real^{Km}$ of system $\Sigma_\rho$, over $k \in \mathcal{K} = \{1,\ldots,K\}$, is the set of all input-output trajectories $(U^{K-1},Y^K)$ satisfying \eqref{1a} over all possible initial conditions $x_\rho(1) \in \Real^{n_\rho}$.
\end{definition}

Hence, classification could be accomplished by identifying to which behaviour $\mathcal{B}_{\rho}$, $\rho \in \mathcal{N}$, the trajectory $(U^{K-1},Y^K)$ belongs. Note, however, that if $\mathcal{B}_{\rho} \cap \mathcal{B}_{\rho^\prime} \neq \emptyset$, for some $\rho,\rho^\prime \in \mathcal{N}$, $\rho \neq \rho^\prime$, trajectories might belong to multiple behaviours, i.e., trajectories in the intersection are not classifiable. This limitation is inherent to the system dynamics and cannot be surpassed by any classifier. In this manuscript, we are interested in forcing the cloud to misclassify trajectories. To induce this, we modify the input-output data that we provide to the cloud so that it appears to have been generated by a different mode. Concretely, given $(U^{K-1},Y^K)$ generated by some mode $\rho \in \mathcal{N}$, and a behaviour $\mathcal{B}_{\rho^\prime}$, $\rho^\prime \in \mathcal{N}$, $\rho \neq \rho^\prime$, we seek a map, $g_{\rho,\rho^\prime}: \mathcal{B}_{\rho} \rightarrow \mathcal{B}_{\rho^\prime}$, referred here as a \emph{distorting map}. That is, the map $(U^{K-1},Y^K) \mapsto g_{\rho,\rho^\prime}(U^{K-1},Y^K)$ takes trajectories from $\mathcal{B}_{\rho}$ and maps them into $\mathcal{B}_{\rho^\prime}$. By passing $(U^{K-1},Y^K)$ through $g_{\rho,\rho^\prime}(\cdot)$ before transmission, we are forcing the cloud to classify $g_{\rho,\rho^\prime}(U^{K-1},Y^K)$ and since $g_{\rho,\rho^\prime}(U^{K-1},Y^K) \in \mathcal{B}_{\rho^\prime}$, the cloud would (ideally) classify the mode as $\rho^\prime$.

\begin{definition}[Distorting Map and Target Mode]
Given two behaviours, $\mathcal{B}_{\rho}$ and $\mathcal{B}_{\rho^\prime}$, $\rho,\rho^\prime \in \mathcal{N}$, $\rho \neq \rho^\prime$, we say that a function $g_{\rho,\rho^\prime}(\cdot)$ is a distorting map if $g_{\rho,\rho^\prime}: \mathcal{B}_{\rho} \rightarrow \mathcal{B}_{\rho^\prime}$. We refer to $\rho^\prime$ as the target mode.
\end{definition}

Note that, because the dynamics of the modes in \eqref{1} is linear, each behaviour $\mathcal{B}_{\rho}$ is an linear subspace. The latter implies that there might exist distorting maps that make the Euclidian distance between $(U^{K-1},Y^K)$ and $g_{\rho,\rho^\prime}(U^{K-1},Y^K)$ arbitrarily large. We do not want to overly distort trajectories. Usually, there is some sensitive information (function of the input-output trajectory $(U^{K-1},Y^K)$) that we would like the cloud to compute accurately. For instance, in intelligent transportation systems, we might want the cloud to accurately compute the average speed of vehicles so that it can send the highway capacity or the shortest route to a destination back to us. To this end, we introduce the notions of \emph{utility} and \emph{utility function} of the trajectory $(U^{K-1},Y^K)$.

\begin{definition}[Utility and Utility Function]
The utility of a trajectory $(U^{K-1},Y^K)$ refers to some sensitive information, denoted as $z(U^{K-1},Y^K) \in \Real^q$, $q \in \Nat$, $z:\Real^{(K-1)l} \times \Real^{Km} \rightarrow \Real^{q}$, the cloud must compute accurately. We refer to the function $z(\cdot)$ as the utility function.
\end{definition}

Then, to maintain the utility of the trajectory after distortion, we require that the utility function evaluated at the distorted data equals the utility of the trajectory $(U^{K-1},Y^K)$. This imposes a constraint on the modes that we can select as target systems and the class of distorting functions that we can use. Concretely, we seek distorting mechanisms, $g_{\rho,\rho^\prime}(\cdot)$, that satisfy $z \circ g_{\rho,\rho^\prime}(U^{K-1},Y^K) = z(U^{K-1},Y^K)$ and map the input-output trajectory $(U^{K-1},Y^K_\rho)$ into the behaviour $\mathcal{B}_{\rho^\prime}$ -- leading to incorrect classification.

In some applications, it might not be realistic to assume that we know the utility function exactly. If $z(\cdot)$ is completely unknown, we would not know how to select $g_{\rho,\rho^\prime}(\cdot)$ to avoid overly distorting the trajectory. So, in the problem formulation introduced above, we are implicitly assuming that $z(\cdot)$ is known. To relax this, we work with utility functions that can be written as the composition of two other functions, an \emph{unknown} function $h:\Real^r \rightarrow \Real^{q}$ and a \emph{known} function $f:\Real^{(K-1)l} \times \Real^{Km} \rightarrow \Real^{r}$, $r \in \Nat$, i.e., $z(\cdot) = h \circ f (\cdot)$. We formulate the problem in terms of the known part of $z(\cdot)$, the function $f (\cdot)$. This is without loss of generality as if the complete $z(\cdot)$ is known, $z(\cdot)=f(\cdot)$ and $h(\cdot)=\text{id}(\cdot)$, where $\text{id}(\cdot)$ denotes the identity map. From a different perspective, some utility functions, even if they are fully known, might be too complicated to work with. Then, \emph{factorising} $z(\cdot)$ as $h \circ f (\cdot)$ and working with a lower complexity function $f(\cdot)$ might make the problem more tractable. Then, if $z(\cdot) = h \circ f(\cdot)$, for some lower (or equal) complexity known function $f(\cdot)$, the aforementioned utility constraint, $z \circ g_{\rho,\rho^\prime} (U^{K-1},Y^K) = z(U^{K-1},Y^K)$, takes the form $h \circ f \circ g_{\rho,\rho^\prime} (U^{K-1},Y^K) = h \circ f(U^{K-1},Y^K)$, which is satisfied if (and only if when $h(\cdot)$ is an injection) $f \circ g_{\rho,\rho^\prime} (U^{K-1},Y^K) = f(U^{K-1},Y^K)$.

There are many practical examples where utility functions can be written as the composition of lower complexity functions. For instance, consider a room with distributed temperature sensors, the utility $z(Y^K)$ could be a binary signal that controls whether the air conditioning must be turn on/off. We do not know how the cloud is actually computing this control signal; however, we do know it is a function of the average temperature in the room over a period of time, $f(Y^K)$. Another example is the computation of total caloric expenditure in fitness smart devices. For each training session, sensors embedded in the device collect data, send it to the cloud, and the cloud returns how many calories we have burned during that training session (this would be the utility $z(Y^K)$). We do not know exactly how they compute the caloric expenditure, but we know this computation depends, e.g., on our average velocity (if running) and maximum/average heart rate during the training session, $f(Y^K)$. Finally, the cost $z(Y^K)$ of residential electricity over a period of time depends on the total power consumption $f(Y^K)$, i.e., $z(Y^K) = h \circ f (Y^K)$, for some unknown $h(\cdot)$.\vspace{1mm}

Next, we formally pose the problem we seek to address.

\begin{problem}[Misclassification-Utility Problem]
Given an input-output trajectory $(U^{K-1},Y^K_\rho)$, a target mode $\rho^\prime$, $\rho,\rho^\prime \in \mathcal{N}$, $\rho \neq \rho^\prime$, and a utility function $z(\cdot) = h \circ f(\cdot)$,\linebreak find a distorting map $g_{\rho,\rho^\prime}: \mathcal{B}_{\rho} \rightarrow \mathcal{B}_{\rho^\prime}$ satisfying:\linebreak $ f \circ g_{\rho,\rho^\prime} (U^{K-1},Y^K_\rho) = f(U^{K-1},Y^K_\rho)$.
\end{problem}

Thus, Problem 1 seeks distorting mechanisms for which the distorted data leads to the same utility as $(U^{K-1},Y^K_\rho)$ -- since $f \circ g_{\rho,\rho^\prime} (U^{K-1},Y^K_\rho) = f(U^{K-1},Y^K_\rho) $ implies $ h \circ f \circ g_{\rho,\rho^\prime} (U^{K-1},Y^K_\rho) = h \circ f(U^{K-1},Y^K_\rho)$ -- and that map the trajectory $(U^{K-1},Y^K_\rho)$ into the behaviour $\mathcal{B}_{\rho^\prime}$.\vspace{.5mm}

Note that, in most of the aforementioned examples of utility functions (and actually in many more not mentioned here), the known part of $z(\cdot)$, $f(\cdot)$, is either an average of some sensor measurements or a weighted sum of them over a period of time, i.e., $f(\cdot)$ \emph{is a linear transformation of the output trajectory $Y^K$}. Motivated by this, we start our analysis considering affine functions for $f(\cdot)$ that only depend on output trajectories $Y^K$. Also, because behaviours are linear subspaces and we consider an affine function $f(\cdot)$, we start considering affine distorting maps $g_{\rho,\rho^\prime}(\cdot)$ too.

\section{Affine Distorting Maps and Utility Functions}

Consider $g_{\rho,\rho^\prime}(\cdot)$ and $f(\cdot)$ of the form:
\begin{subequations}\label{affine_maps}
\begin{align}
g_{\rho,\rho^\prime}(Y,U) &\coloneqq \begin{pmatrix} Y + \Delta_Y \\ U + \Delta_U \end{pmatrix},\label{map1}\\
f(Y) &\coloneqq FY + \mu,\label{map2}
\end{align}
\end{subequations}
with $Y,\Delta_Y \in \Real^{Km}$, $U,\Delta_U \in \Real^{(K-1)l}$, $\Delta \coloneqq \text{col}[\Delta_Y,\Delta_U]$, $F \in \Real^{q \times Km}$, and $\mu \in \Real^q$. It follows that Problem 1 amounts to finding $\Delta \in \Real^{K(m+l)-l}$ such that $\text{col}[U^{K-1} + \Delta_U, Y^K_\rho + \Delta_Y] \in \mathcal{B}_{\rho^\prime}$, $\rho,\rho^\prime \in \mathcal{N}$, $\rho \neq \rho^\prime$, for some target mode $\rho^\prime$, and $F(Y^K_\rho + \Delta_Y) + \mu = FY^K_\rho + \mu$ (i.e., $\Delta_Y \in \text{Ker}[F]$).

If the complete input-output trajectory $(U^{K-1},Y^K_\rho)$ is available before transmission, i.e., all data is collected before it is sent to the cloud, the problem of finding $\Delta$ solving Problem 1 (for the setting described above) is purely algebraic. That is, one might lift the system dynamics for the length of the trajectory and pose the problem of finding $\Delta$ as the solution of some linear equations. However, in most real-time applications, input-output data is sent to the cloud immediately after it is generated. Thus, a more realistic configuration is to modify $(u(k),y(k))$ recursively and in real-time so that the modified data, say $(\bar{u}(k),\bar{y}(k))$, $k \in \mathcal{K}$, satisfies $F\bar{Y}^K = F Y^K$ (same utility) and $(\bar{U}^{K-1},\bar{Y}^K) \in \mathcal{B}_{\rho^\prime}$ (belongs to the target mode behaviour), for some target mode $\rho^\prime \in \mathcal{N}$, $\bar{Y}^K = \text{col}[\bar{y}(1),\bar{y}(2),\ldots,\bar{y}(K)]$, and $\bar{U}^{K-1} = \text{col}[\bar{u}(1),\bar{u}(2),\ldots,\bar{u}(K-1)]$.

The target mode dynamics, $\Sigma_{\rho^\prime}$, is characterized by the triple $(A_{\rho^\prime},B_{\rho^\prime},C_{\rho^\prime})$, $\rho^\prime \in \mathcal{N}$, as introduced in \eqref{1}. By Definition 1, any input-output trajectory $(\bar{U}^{K-1},\bar{Y}^K)$ in $\mathcal{B}_{\rho^\prime}$ satisfies the difference equations:
\begin{align}
&\left\{ \begin{array}{l}
\bar{x}(k+1) = A_{\rho^\prime} \bar{x}(k) + B_{\rho^\prime} \bar{u}(k),\\[0.5mm]
\hspace{6mm}\bar{y}(k) = C_{\rho^\prime} \bar{x}(k),
\end{array} \right. \label{target}
\end{align}
for some initial condition $\bar{x}(1) \in \Real^{n_{\rho^\prime}}$. Therefore, we can generate trajectories from $\mathcal{B}_{\rho^\prime}$ by fixing the input sequence $\bar{u}(k) \in \Real^l$, $k \in \{1,2,\ldots,K-1\}$, and passing it through \eqref{target} for some initial condition, to obtain an output sequence $\bar{y}(k) \in \Real^m$, $k \in \mathcal{K}$. By construction, the corresponding trajectory $(\bar{U}^{K-1},\bar{Y}^K)$ belongs to $\mathcal{B}_{\rho^\prime}$. Thus, we can use \eqref{target} to recursively generate trajectories from the target mode behaviour. The idea is that if we send these trajectories through the network (instead of the actual $(U^{K-1},Y^K)$), the cloud would classify the mode as $\rho^\prime$. However, we cannot just send any trajectory. We need $Y^K$ and $\bar{Y}^K$ to lead to the same utility, i.e., $F\bar{Y}^K = FY^K$ (see \eqref{affine_maps}). Note that if $\bar{Y}^K = Y^K + \Delta_Y$, $F\bar{Y}^K = FY^K$, if and only if $\Delta_Y \in \text{Ker}[F]$. Let $\Delta_Y = \text{col}[\delta_Y(1),\ldots,\delta_Y(K)]$, $\delta_Y(i) \in \Real^m$, $i \in \mathcal{K}$; then, $\bar{Y}^K = Y^K + \Delta_Y$ can be written as $\bar{y}(k) = y(k) + \delta_Y(k)$, $k \in \mathcal{K}$. Hence, we can address Problem 1 recursively and in real-time by designing an artificial input sequence $\bar{u}(k)$, $k \in \{1,2,\ldots,K-1\}$, and an initial condition $\bar{x}(1) \in \Real^{n_{\rho^\prime}}$ such that $\bar{y}(k)$ in \eqref{target} satisfies $\bar{y}(k) = y(k) + \delta_Y(k)$ for some $\Delta_Y \in \text{Ker}[F]$.

Let $\bar{u}(k) = \bar{u}^1(k) + \bar{u}^2(k)$, and write the state, $\bar{x}(k)$, and output, $\bar{y}(k)$, of \eqref{target} as $\bar{x}(k) = \bar{x}^1(k) + \bar{x}^2(k)$ and $\bar{y}(k) = \bar{y}^1(k) + \bar{y}^2(k)$, where $(\bar{x}^1(k),\bar{y}^1(k))$ denotes the part of $(\bar{x}(k),\bar{y}(k))$ driven by $\bar{u}^1(k)$ and $(\bar{x}^2(k),\bar{y}^2(k))$ the part driven by $\bar{u}^2(k)$. Using this new notation and superposition of linear systems, we can write \eqref{target} as
\begin{subequations}
\begin{align}
&\left\{ \begin{array}{l}
\bar{x}^1(k+1) = A_{\rho^\prime} \bar{x}^1(k) + B_{\rho^\prime} \bar{u}^1(k),\\[0.5mm]
\hspace{6mm}\bar{y}^1(k) = C_{\rho^\prime} \bar{x}^1(k),
\end{array} \right. \label{target1} \\
&\left\{ \begin{array}{l}
\bar{x}^2(k+1) = A_{\rho^\prime} \bar{x}^2(k) + B_{\rho^\prime} \bar{u}^2(k),\\[0.5mm]
\hspace{6mm}\bar{y}^2(k) = C_{\rho^\prime} \bar{x}^2(k),
\end{array} \right. \label{target2}\\[1mm]
&\hspace{12.5mm} \bar{y}(k) = \bar{y}^1(k) + \bar{y}^2(k),\label{target3}
\end{align}
\end{subequations}
with corresponding initial conditions $\bar{x}^1(1),\bar{x}^2(1) \in \Real^n$ satisfying $\bar{x}(1) = \bar{x}^1(1) + \bar{x}^2(1)$. Then, an approach to enforce $\bar{y}(k) = y(k) + \delta_Y(k)$, $k \in \mathcal{K}$, is to design $\bar{u}^1(k)$ in \eqref{target1} such that $\bar{y}^1(k) = C_{\rho^\prime} \bar{x}^1(k) = y(k)$ (\emph{output regulation}), $k \in \mathcal{K}$, and $u^2(k)$ in \eqref{target2} to enforce $\bar{y}^2(k) = C_{\rho^\prime} \bar{x}^2(k) = \delta_Y(k)$ (\emph{utility invariance}), $k \in \mathcal{K}$, and apply the combined $\bar{u}(k) = \bar{u}^1(k) + \bar{u}^2(k)$ to the virtual target system \eqref{target}. By construction, the resulting $\bar{y}(k)$ satisfies $\bar{y}(k) = y(k) + \delta_Y(k)$ and the input-output trajectory $(\bar{U}^{K-1},\bar{Y}^K)$ belongs to $\mathcal{B}_{\rho^\prime}$.

Using output-regulation techniques \cite{Francis_Bruce}-\cite{FRANCIS1976457}, we design input $\bar{u}^1(k)$ and the initial condition $\bar{x}^1(1)$ to regulate the error, $r(k) \coloneqq \bar{y}^1(k)-y(k)$, given $(u(k),y(k))$, the true mode dynamics $\Sigma_\rho$, and the target mode $\rho^\prime$. Input $u^2(k)$ is used to steer $\bar{y}^2(k)$, $k \in \mathcal{K}$, to an element, $\Delta_Y$, in the kernel of $F$. Since we know $F$ a priori (before starting the system operation), we can design $u^2(k)$ off-line, i.e., without using real-time data $(u(k),y(k))$. In particular, we lift the target system dynamics \eqref{target2} over $k \in \mathcal{K}$ and cast the problem of finding $\bar{x}^2(1)$ and $\bar{u}^2(k)$, $k \in \{1,\ldots,K-1\}$, in terms of the solution of some linear equations.

\subsection{Virtual Output Regulation}

Consider the trajectory $(U^{K-1},Y^K)$ generated in real-time by system $\Sigma_\rho$, $\rho \in \Nat$. At every $k \in \mathcal{K}$, the input-output data available to design $\bar{u}^1(k)$ is $(U^{k-1},Y^k)$. For ease of presentation, we assume that the state $x_\rho(k)$ of system $\Sigma_\rho$ is available for feedback. However, when this is not true, given observability of $(A_\rho,C_\rho)$, we can recover the state $x_\rho(k)$ from $(U^{k-1},Y^k)$ after $n$ time-steps (note that, in general, $n \ll K$). In this case, we would have to wait for $n$ time-steps before we start sending the corrupted data to the cloud. An alternative would be to use \emph{internal model principle} techniques to synthesize \emph{dynamic regulators} \cite{FRANCIS1976457} for $\bar{u}^1(k)$. In this manuscript, however, we assume that $x_\rho(k)$ is available at every $k \in \mathcal{K}$ and work with \emph{static regulators} to enforce $\bar{y}^1(k) = y(k)$, $k \in \mathcal{K}$.

Consider the following state controller for system \eqref{target1}:
\begin{align}\label{controller1}
\bar{u}^1(k) = R\bar{x}^1(k) + Lx_\rho(k) + Su(k),
\end{align}
with true system state $x_\rho(k) \in \Real^{n_\rho}$ and input $u(k) \in \Real^l$ of \eqref{1}, virtual state $\bar{x}^1(k) \in \Real^{n_{\rho^\prime}}$ of \eqref{target1}, and matrices $R \in \Real^{l \times n_{\rho^\prime}}$, $L  \in \Real^{l \times n_{\rho}}$, and $S  \in \Real^{l \times l}$. The feedback term, $R\bar{x}^1(k)$, is used to enforce internal stability of \eqref{target1}-\eqref{controller1} only, i.e., matrix $R$ is selected so that $(A_{\rho^\prime}+B_{\rho^\prime}R)$ is Schur stable. Such an $R$ always exists due to controllability of $(A_{\rho^\prime},B_{\rho^\prime})$. We need internal stability to prevent $\bar{u}^1(k)$ from growing unbounded. The remaining terms in \eqref{controller1}, $Lx_\rho(k)$ and $Su(k)$, are used to enforce $r(k) = \bar{y}^1(k)-y(k) = \mathbf{0}$, for all $u(k) \in \Real^l$, $x_\rho(k) \in \Real^{n_\rho}$, and $k \in \mathcal{K}$.

\begin{problem}[Virtual Output Regulation]\label{problem_2}
Given input-output real-time data $(u(k),y(k))$ generated by mode $\Sigma_\rho$, the target mode dynamics \eqref{target1}, controller \eqref{controller1}, and a matrix $R$ so that $(A_{\rho^\prime}+B_{\rho^\prime}R)$ is Schur, find (if possible) matrices $(L,S)$ in \eqref{controller1} and initial condition $\bar{x}^1(1)$ of the virtual system \eqref{target1} such that $\bar{y}^1(k) = y(k)$ for all $u(k) \in \Real^l$, $x_\rho(k) \in \Real^{n_\rho}$, and $k \in \mathcal{K}$.
\end{problem}

\begin{theorem}\label{theorem1}
Problem \ref{problem_2} is solvable if and only if there exist matrices $\Pi \in \Real^{n_{\rho^\prime} \times n_\rho}$, $\Gamma \in \Real^{l \times n_{\rho}}$, and $\Theta \in \Real^{l \times l}$ that are a solution to the regulator equations:
\begin{align}\label{reg}
\left\{ \begin{array}{l}
A_{\rho^\prime} \Pi - \Pi A_{\rho} + B_{\rho^\prime}\Gamma = \mathbf{0},\\
C_{\rho^\prime} \Pi - C_{\rho} = \mathbf{0},\\
B_{\rho^\prime} \Theta - \Pi B_{\rho} = \mathbf{0}.
\end{array} \right.
\end{align}
\end{theorem}
\textbf{\emph{Proof.}} \emph{Sufficiency:} Let \eqref{reg} be satisfied for some $(\Pi,\Gamma,\Theta)$, define $e(k) := \bar{x}^1(k) - \Pi x_\rho(k)$, and consider the closed-loop dynamics \eqref{1},\eqref{target1},\eqref{controller1}. Then, $e(k+1)$ and the regulation error, $r(k) = \bar{y}^1(k) - y(k)$, can be written as
\begin{align}\label{reg1}
e(k+1) &= (A_{\rho^\prime}+B_{\rho^\prime}R)e(k) + \big( B_{\rho^\prime} S - \Pi B_{\rho} \big)u(k)\\
&+ \big(  ( A_{\rho^\prime} + B_{\rho^\prime}R )\Pi - \Pi A_{\rho} + B_{\rho^\prime}L \big)x_\rho(k),\notag\\
r(k) &= (C_{\rho^\prime} \Pi - C_{\rho})x_\rho(k) + C_{\rho^\prime}e(k).\label{reg2}
\end{align}
Let $L = \Gamma - R\Pi$, $S = \Theta$, and $\bar{x}^1(1) = \Pi x_\rho(1)$. Then, by \eqref{reg}, $e(k+1) = (A_{\rho^\prime}+B_{\rho^\prime}R)e(k)$ and $r(k) = C_{\rho^\prime}e(k)$. Moreover, $e(1) = \bar{x}(1) - \Pi x_\rho(1) = \mathbf{0}$ (because $\bar{x}^1(1) = \Pi x_\rho(1)$). It follows that $e(k) = \mathbf{0}$ for $k \in \mathcal{K}$ and hence $r(k) = \bar{y}^1(k) - y(k) = \mathbf{0}$ for all $k \in \mathcal{K}$. \emph{Necessity:} Let $r(k) = \bar{y}^1(k) - y(k) = C_{\rho^\prime}\bar{x}^1(k) - C_\rho x_\rho(k)  = \mathbf{0}$, for all $u(k) \in \Real^l$, $x_\rho(k) \in \Real^{n_\rho}$, and $k \in \mathcal{K}$. Consider $e(k) = \bar{x}^1(k) - \Pi x_\rho(k)$, for some arbitrary $\Pi \in \Real^{n_{\rho^\prime} \times n_\rho}$, and note that the transformation $(x_\rho(k),\bar{x}^1(k)) \rightarrow (x_\rho(k),e(k))$ is invertible for any $\Pi$. It follows that $r(k)$ can always be written as in \eqref{reg2} and because $r(k) = \mathbf{0}$, $(C_{\rho^\prime} \Pi - C_{\rho})x_\rho(k) + C_{\rho^\prime}e(k) = \mathbf{0}$. The latter is satisfied for all $x_\rho(k) \in \Real^{n_\rho}$ (including $x_\rho(k) \in \text{Im}[C_{\rho^\prime} \Pi - C_{\rho}]$) and $u(k) \in \Real^l$; hence, necessarily, $C_{\rho^\prime} \Pi - C_{\rho} = \mathbf{0}$ (i.e., the second equality in \eqref{reg} is satisfied for some $\Pi$), and $e(k) \in \text{Ker}[C_{\rho^\prime}]$. Consider now $e(k+1)$ (given by \eqref{reg1}), and note that $e(k+1) \in \text{Ker}[C_{\rho^\prime}]$ for all $x_\rho(k)$ and $u(k)$ in $\Real^{n_\rho}$ and $\Real^l$, respectively, if and only if $e(k) = \mathbf{0}$ for all $k \in \mathcal{K}$. The latter implies that $( A_{\rho^\prime} + B_{\rho^\prime}R )\Pi + B_{\rho^\prime}L = \Pi A_{\rho}$ and $B_{\rho^\prime} S = \Pi B_{\rho}$, which equal the first and third equations in \eqref{reg} with $\Gamma = L + R\Pi$ and $\Theta = S$. \hfill $\blacksquare$

\begin{corollary}\label{corollary1}
Consider Problem \ref{problem_2} and let $\Pi \in \Real^{n_{\rho^\prime} \times n_\rho}$, $\Gamma \in \Real^{l \times n_{\rho}}$, and $\Theta \in \Real^{l \times l}$ solve the regulator equations \eqref{reg}. Then, $L = \Gamma - R\Pi$, $S = \Theta$, and $\bar{x}^1(1) = \Pi x_\rho(1)$ are a solution to Problem \ref{problem_2}.
\end{corollary}
\textbf{\emph{Proof.}} Corollary \ref{corollary1} follows from the sufficiency part of the proof of Theorem \ref{theorem1}. \hfill $\blacksquare$

Theorem \ref{theorem1} provides necessary and sufficient conditions for Problem \ref{problem_2} to have a solution in terms of the solution of the regulator equations \eqref{reg}. Once we have a solution (this solution might not be unique), for given $R$ so that $(A_{\rho^\prime}+B_{\rho^\prime}R)$ is Schur, we can compute matrices $(L,S)$ and initial condition $\bar{x}^1(1)$ to realize the controller $\bar{u}^1(k)$ in \eqref{controller1} using Corollary \ref{corollary1}.

\begin{remark}
Note that any controller $\bar{u}^1(k)$ in \eqref{controller1} and initial condition $\bar{x}^1(1)$ of \eqref{target1} solving Problem \ref{problem_2} provide already a solution to Problem 1 (for the class of $f(\cdot)$ and $g_{\rho,\rho^\prime}(\cdot)$ introduced above). That is, these $\bar{u}^1(k)$ and $\bar{y}^1(k)$, $k \in \mathcal{K}$, belong to $\mathcal{B}_{\rho^\prime}$ and, because $\bar{y}^1(k) = y(k)$, $Y^K$ and $\bar{Y}^K = \text{\emph{col}}[\bar{y}^1(1),\ldots,\bar{y}^1(k)]$ have the same utility, i.e., $FY^K = F\bar{Y}^K$. However, if we share $\bar{u}^1(k)$ and $\bar{y}^1(k)$, the cloud would still get the true output data -- with a different input sequence though. In the next subsection, we provide tools for properly distorting output data to avoid sharing the true output sequence with the cloud.
\end{remark}

\subsection{Utility Invariance (Batch Approach)}

In this subsection, we provide tools for designing $\bar{u}^2(k)$ and $\bar{x}^2(1)$ in \eqref{target2} so that $\bar{y}^2(k)$, $k \in \mathcal{K}$, is steered to an element, $\Delta_Y \neq \mathbf{0}$, in the kernel of $F$. We use the resulting $\bar{u}^2(k)$ and a $\bar{u}^1(k)$ synthesized using Corollary \ref{theorem1} to construct the actual input $\bar{u}(k) = \bar{u}^1(k) + \bar{u}^2(k)$ and initial condition $\bar{x}(1) = \bar{x}^1(1) + \bar{x}^2(1)$ for the virtual system \eqref{target}. Because we know $F$ a priori (before starting the system operation) and $\bar{u}^1(k)$ is already being used to handle real-time data, we can actually design $u^2(k)$ off-line, i.e., independent of $(u(k),y(k))$. To do so, we lift the target system dynamics \eqref{target2}, over $k \in \mathcal{K}$, and cast the problem of finding $\bar{x}^2(1)$ and $\bar{u}^2(k)$, $k \in \{1,\ldots,K-1\}$, in terms of the solution of some linear equations.

We aim at enforcing that the sequence of virtual outputs $\bar{y}^2(k)$, $k \in \mathcal{K}$, is contained in the kernel of $F$. If $\text{Ker}[F]$ is trivial, the only vector to which we can drive $\bar{y}^2(k)$ is the zero vector. In this case, $\bar{u}^1(k)$ solving Problem \ref{problem_2} and $\bar{y}^1(k) = y(k)$ are the only option for solving Problem 1 (see Remark 1). Therefore, a necessary condition for Problem 1 to have a different solution ($\bar{y}(k) \neq y(k)$) is that $F$ has a nontrivial kernel.

\begin{assumption}
The kernel of $F \in \Real^{q \times Km}$ is nontrivial.
\end{assumption}

Consider $\tilde{Y}^K :=\text{col}[\bar{y}^2(1),\ldots,\bar{y}^2(K)]$. The stacked vector $\tilde{Y}^K$ can be written explicitly in terms of $\bar{x}^2(1)$ and $\tilde{U}^{K-1} :=\text{col}[\bar{u}^2(1),\ldots,\bar{u}^2(K-1)]$ as follows
\begin{equation}\label{stackedY}
\begingroup
\renewcommand*{\arraycolsep}{1pt}
\left\{
\begin{aligned}
  &\tilde{Y}^K = \mathcal{O}_K\bar{x}^2(1) + \mathcal{T}_K\tilde{U}^{K-1},\\
  &\mathcal{T}_K := \begin{bmatrix} \mathbf{0} & \mathbf{0} & \cdots & \mathbf{0} \\ C_{\rho^\prime}B_{\rho^\prime} & \mathbf{0} & \cdots & \mathbf{0} \\ C_{\rho^\prime}A_{\rho^\prime}B_{\rho^\prime} & C_{\rho^\prime}B_{\rho^\prime} & \cdots & \mathbf{0} \\ \vdots & \vdots & \ddots & \vdots \\[1mm] C_{\rho^\prime}(A_{\rho^\prime})^{K-2}B_{\rho^\prime} & C_{\rho^\prime}(A_{\rho^\prime})^{K-3}B_{\rho^\prime} & \cdots & C_{\rho^\prime}B_{\rho^\prime}  \end{bmatrix},\\
  &\mathcal{O}_K := \begin{bmatrix} C_{\rho^\prime} \\ C_{\rho^\prime}A_{\rho^\prime} \\ \vdots \\ C_{\rho^\prime}(A_{\rho^\prime})^{K-1}  \end{bmatrix},
\end{aligned}
\right. \endgroup
\end{equation}

\begin{problem}[Utility Invariance]\label{problem_3}
Given the utility function \eqref{map2} and the target mode dynamics \eqref{target2}, find (if possible) an initial condition $\bar{x}^2(1) \in \Real^{n_{\rho^\prime}}$ of \eqref{target2} and a sequence of inputs $\tilde{U}^{K-1} =\text{\emph{col}}[\bar{u}^2(1),\ldots,\bar{u}^2(K-1)]$ such that $\mathcal{O}_K\bar{x}^2(1) + \mathcal{T}_K\tilde{U}^{K-1} \in \text{\emph{Ker}}[F]$, with $\mathcal{O}_K$ and $\mathcal{T}_K$ as defined in \eqref{stackedY}.
\end{problem}

\begin{lemma}\label{lemma1}
Consider the target mode behaviour $\mathcal{B}_{\rho^\prime}$. Problem \ref{problem_3} is solvable if and only if $\mathcal{B}_{\rho^\prime} \cap \text{\emph{Ker}}[F] \neq \emptyset$.
\end{lemma}
\textbf{\emph{Proof.}} Note that $\mathcal{B}_{\rho^\prime}$ can be written as $\mathcal{B}_{\rho^\prime} = \{ \mathcal{O}_K x + \mathcal{T}_K U | x \in \Real^{n_{\rho^\prime}},U \in \Real^{(K-1)l} \}$. Then, $\mathcal{B}_{\rho^\prime} \cap \text{Ker}[F] \neq \emptyset$ implies that there exists some $x^*$ and $U^*$ such that $\mathcal{O}_Kx^* + \mathcal{T}_KU^* \in \text{Ker}[F]$. Conversely, if for some $(x^*,U^*)$ $\mathcal{O}_Kx^* + \mathcal{T}_KU^* \in \text{Ker}[F]$, there is at least one vector in $\mathcal{B}_{\rho^\prime}$, $\mathcal{O}_Kx^* + \mathcal{T}_KU^*$, that is contained in $\text{Ker}[F]$. \hfill $\blacksquare$

\begin{theorem}\label{theorem2}
Problem \ref{problem_3} is solvable if and only if there exist vectors $x \in \Real^{n_{\rho^\prime}}$, $U \in \Real^{(K-1)l}$, and $\theta \in \Real^{Km}$ solution to the linear equations: \vspace{2mm}
\begin{align}\label{map3}
\Bigg( \mathcal{O}_K \hspace{3mm} \mathcal{T}_K \hspace{3mm} (F^+F - I_{Km}) \Bigg)
\begin{pmatrix}
  x \\ U \\ \theta
\end{pmatrix} = \mathbf{0}.
\end{align}
\end{theorem}
\textbf{\emph{Proof.}} Matrix $(I_{Km} - F^+F)$ is a basis of the right kernel of $F$. It follows that $(I_{Km} - F^+F)\theta \in \text{Ker}[F]$ for all $\theta \in \Real^{Km}$. Let $(x,U,\theta)$ solve \eqref{map3}. Then, $\mathcal{O}_K x + \mathcal{T}_K U = (I_{Km} - F^+F)\theta \in \text{Ker}[F]$. Conversely, let $(x,U)$ satisfy $\mathcal{O}_K x + \mathcal{T}_K U \in \text{Ker}[F]$. Then, because $(I_{Km} - F^+F)$ is a basis of the kernel of $F$, there most exists $\theta \in \Real^{Km}$ so that $\mathcal{O}_K x + \mathcal{T}_K U = (I_{Km} - F^+F)\theta$. \hfill $\blacksquare$

\begin{corollary}\label{corollary2}
Consider Problem \ref{problem_3} and let $x \in \Real^{n_{\rho^\prime}}$, $U \in \Real^{(K-1)l}$, and $\theta \in \Real^{Km}$ solve \eqref{map3}. Then, $\bar{x}^2(1) = x$ and $\tilde{U}^{K-1} = U$ are a solution to Problem \ref{problem_3}.
\end{corollary}
\textbf{\emph{Proof.}} Corollary \ref{corollary2} follows from the proofs of Lemma 1 and Theorem 2. \hfill $\blacksquare$ \vspace{2mm}

Theorem \ref{theorem2} provides necessary and sufficient conditions for Problem \ref{problem_3} to have a solution in terms of the solution of \eqref{map3}. If there exists a solution, using Corollary \ref{corollary1}, we can compute the initial condition $\bar{x}^2(1)$ and the sequence of controllers $\tilde{U}^{K-1} =\text{col}[\bar{u}^2(1),\ldots,\bar{u}^2(K-1)]$ to drive system \eqref{target2} so that $\tilde{Y}^K  \in \text{Ker}[F]$.

\newpage
\begin{remark}
For given mode $\rho^\prime \in \mathcal{N}$ and utility matrix $F$, there either do not exist solutions to \eqref{map3} or the solution is unique or there exist an infinite number of solutions. In the latter case, the set of solutions form a linear subspace, which implies that $\tilde{Y}^K$ can be chosen arbitrarily large. This is the most appealing case to us as we can induce arbitrarily large distortion without affecting the utility of the trajectory.
\end{remark}

In the next subsection, we provide a synthesis procedure to summarize the results presented above.

\subsection{Synthesis Procedure:}\label{synth}

\noindent\rule{\hsize}{1pt}\vspace{.2mm}
\textbf{Synthesis:}\\[.5mm]
\textbf{1)} Given the mode dynamics $\Sigma_\rho$ in \eqref{1}, $\rho \in \mathcal{N}$, that will generate the input-output trajectory, select a target mode $\Sigma_{\rho^\prime}$, $\rho^\prime \in \mathcal{N}$, $\rho \neq \rho^\prime$.\\[.5mm]
\textbf{2)} Using the true system matrices $(A_\rho,B_\rho,C_\rho)$ and the target mode matrices $(A_{\rho^\prime},B_{\rho^\prime},C_{\rho^\prime})$, seek a solution $\Pi \in \Real^{n_{\rho^\prime} \times n_\rho}$, $\Gamma \in \Real^{l \times n_{\rho}}$, and $\Theta \in \Real^{l \times l}$ to the regulator equations \eqref{reg}.\\[.5mm]
\textbf{3)} Select any matrix $R$ so that $(A_{\rho^\prime}+B_{\rho^\prime}R)$ is Schur, and compute matrices $(L,S)$ of controller $\bar{u}^1(k)$ in \eqref{controller1} and the initial condition $\bar{x}^1(1)$ of \eqref{target1} using Corollary 1.\\[.5mm]
\textbf{4)} Consider the virtual target system  \eqref{target1}, with initial $\bar{x}^1(1)$, and close it with controller $\bar{u}^1(k)$ in \eqref{controller1}.\\[.5mm]
\textbf{5)} Given the trajectory length $K \in \Nat$ and the utility matrix $F$ in \eqref{map2}, compute matrices $\mathcal{O}_K$ and $\mathcal{T}_K$ in \eqref{stackedY} and seek a solution $x \in \Real^{n_{\rho^\prime}}$, $U \in \Real^{(K-1)l}$, and $\theta \in \Real^{Km}$ to the linear equations \eqref{map3}.\\[.5mm]
\textbf{6)} Compute the initial condition $\bar{x}^2(1)$ of \eqref{target2} and the sequence of controllers $\{\bar{u}^2(1),\ldots,\bar{u}^2(K-1)\}$ using Corollary \ref{corollary2}.\\[.5mm]
\textbf{7)} Consider the virtual target system  \eqref{target2}, with initial $\bar{x}^2(1)$, and close it with controller $\bar{u}^2(k)$, $k \in \mathcal{K}$.\\[.5mm]
\textbf{8)} Compute the combined input $\bar{u}(k) = \bar{u}^1(k) + \bar{u}^2(k)$ and corresponding output $\bar{y}(k) = \bar{y}^1(k) + \bar{y}^2(k)$ in \eqref{target3}, and send $(\bar{u}(k),\bar{y}(k))$ to the cloud in real-time.\\
\vspace{.2mm}\noindent\rule{\hsize}{1pt}

\section{Academic Example}

\begin{figure}[t]
  \centering
  \includegraphics[scale=.15]{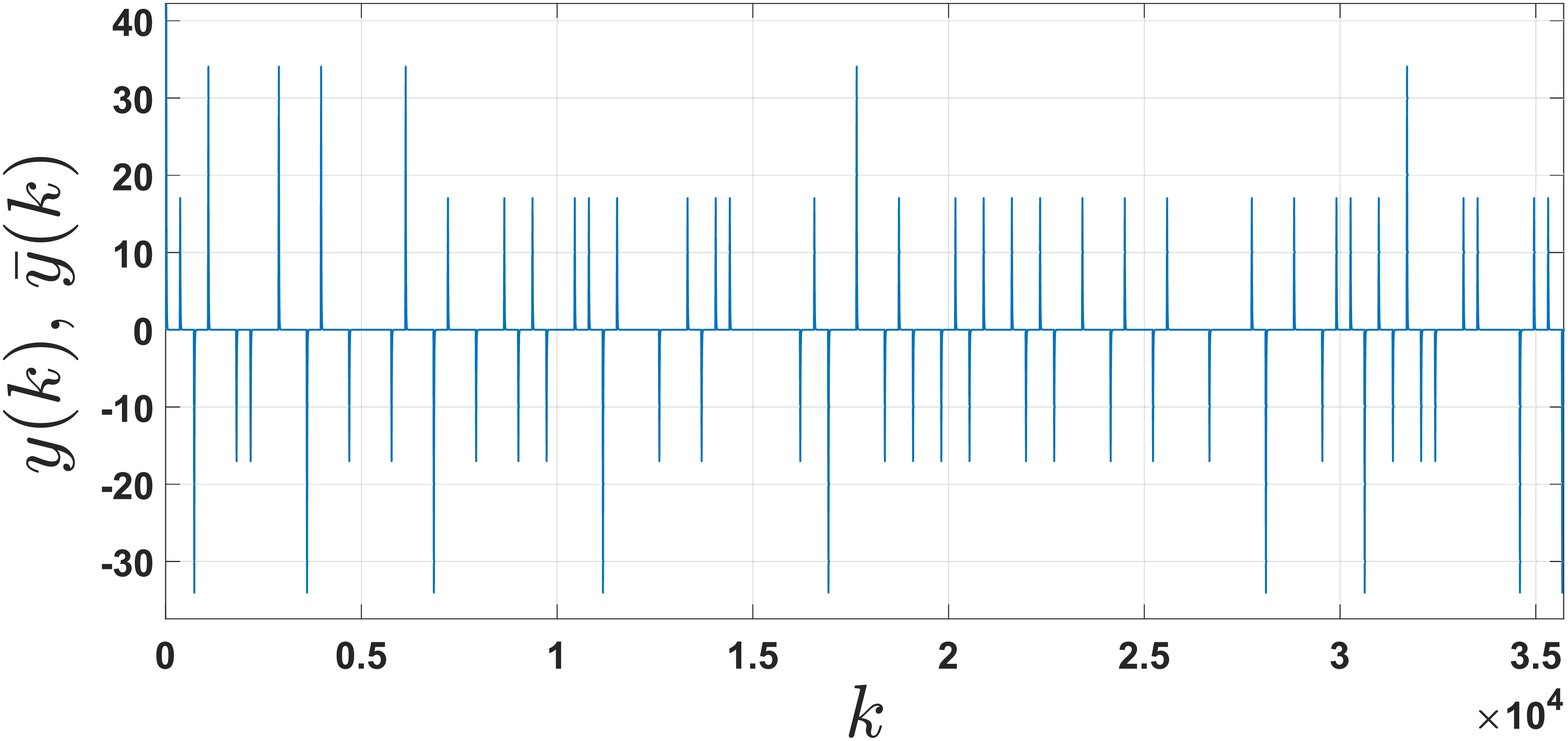}
  \caption{Traces of $y(k)$ and $\bar{y}^1(k)$ (they are indistinguishable). }\label{fig1}
  \vspace*{\floatsep}
  \centering
  \includegraphics[scale=.15]{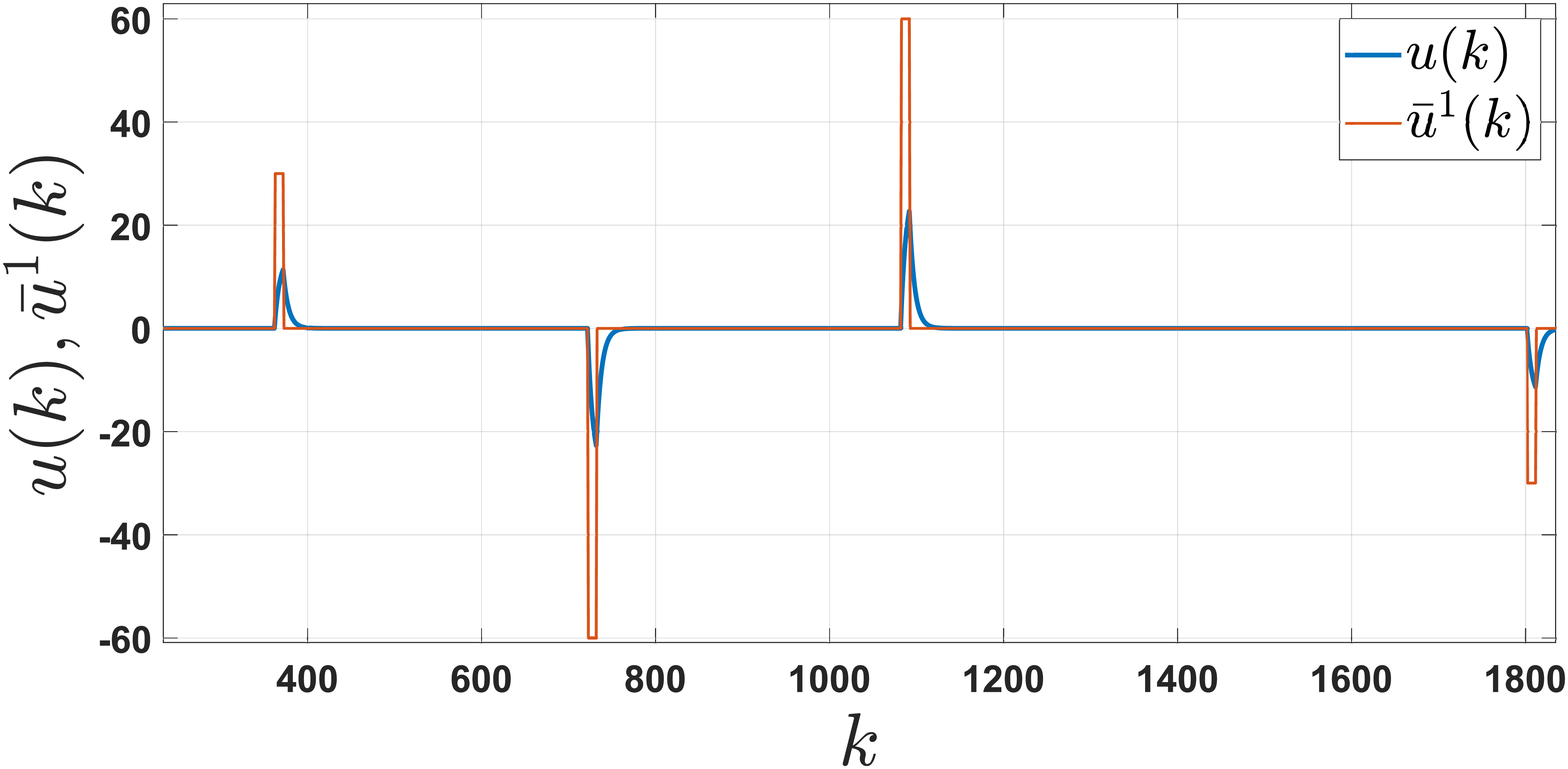}
  \caption{Traces of $u(k)$ and $\bar{u}^1(k)$. }\label{fig2}
\end{figure}

\begin{figure}[t]
  \centering
  \includegraphics[scale=.15]{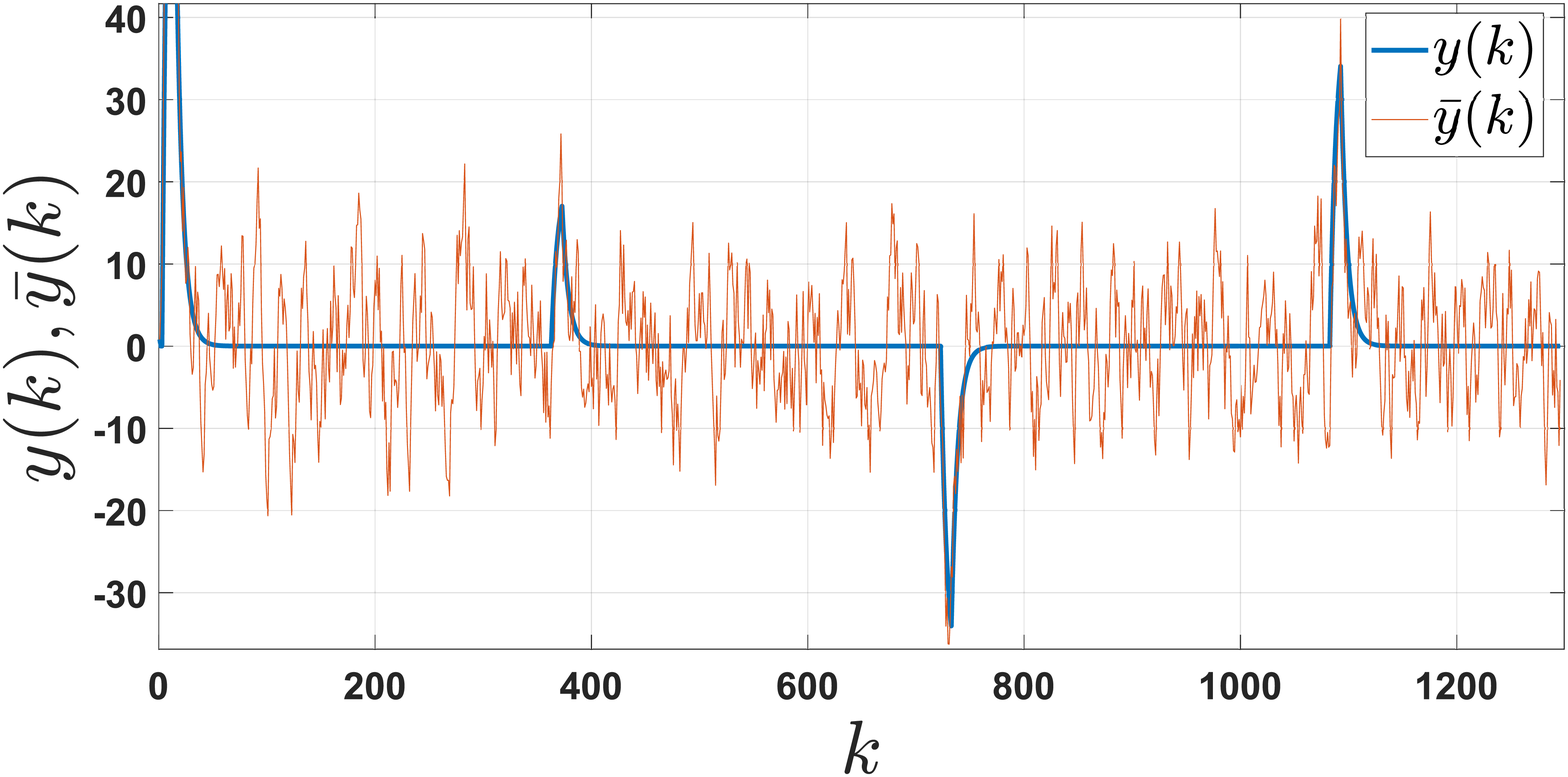}
  \caption{Traces of $y(k)$ and $\bar{y}(k)$. }\label{fig3}
  \vspace*{\floatsep}
  \includegraphics[scale=.15]{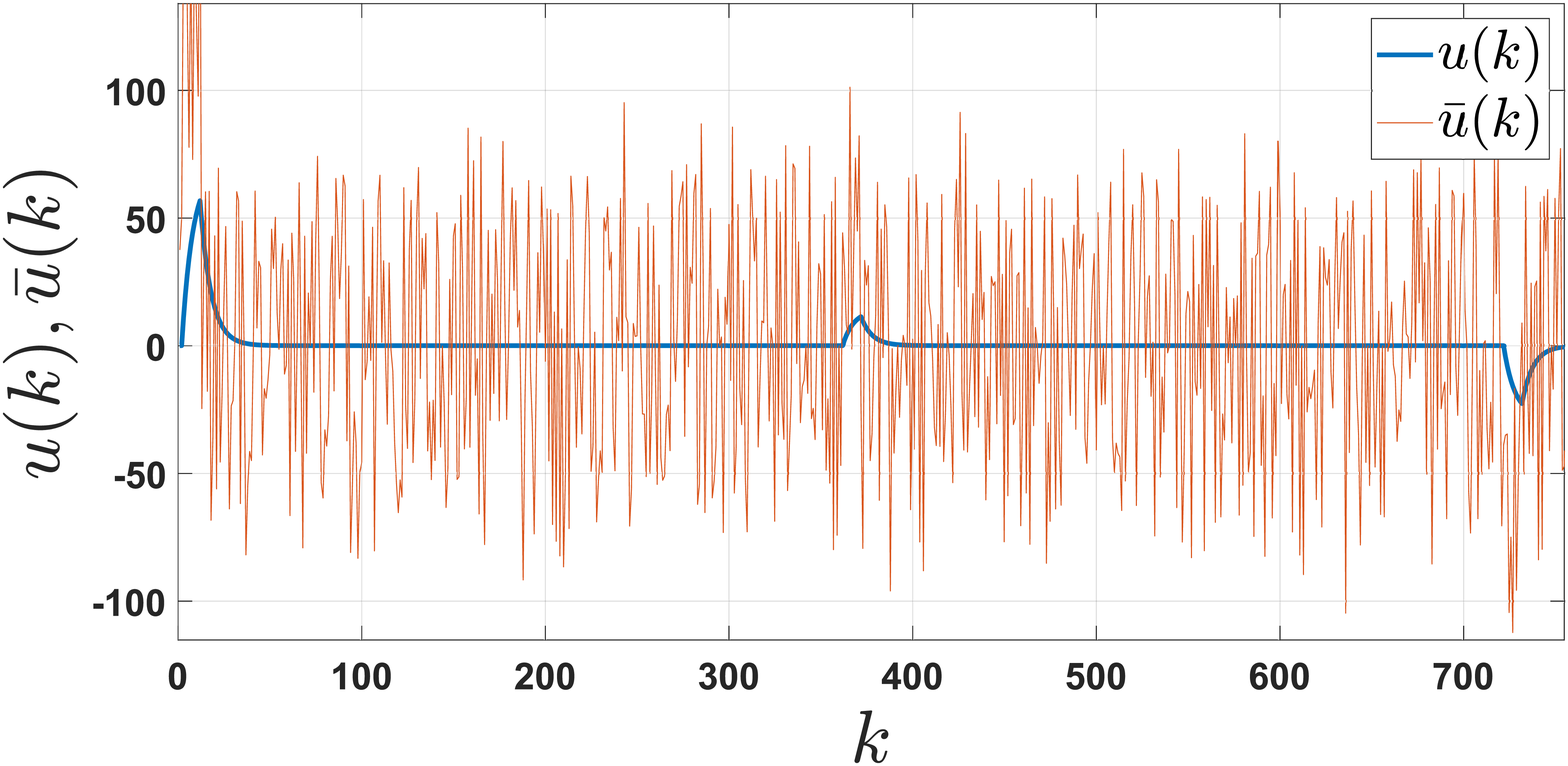}
  \caption{Traces of $u(k)$ and $\bar{u}(k)$. }\label{fig4}
\end{figure}

Consider the following third order continuous-time longitudinal vehicle dynamics:\vspace{.5mm}
\begin{equation}\label{exp_1}
\left\{
\begin{array}{l}
  \dot{q}(t) = v(t),\\[0.5mm]
  \dot{v}(t) = a(t), \\[0.5mm]
  \dot{a}(t) = -\dfrac{1}{\tau}a(t) + \dfrac{\beta}{\tau}u(t),\\[2mm]
       y(t)  = a(t),
\end{array}
\right. \vspace{.5mm}
\end{equation}
where $q \in \Real$, $v \in \Real$, and $a \in \Real$ denote, respectively, position, velocity, and acceleration of the vehicle, $\tau \in \Real_{>0}$ is the inertia time-lag in the power train, and $\beta \in \Real_{>0}$ denotes the engine performance coefficient. We have nominal engine performance with $\beta=1$, decreased performance with $\beta<1$, due to, e.g., wear or a poor aerodynamic design, and increased performance with $\beta>1$, due to spoilers and high aerodynamic efficiency. This model has been extensively used in vehicle platooning research, see, e.g., \cite{Ploega}-\nocite{Ploegb}\cite{Harfouch} and references therein. We simulate trajectories of an expensive sports car using \eqref{exp_1} with $\tau = 0.01$ and $\beta = 1.50$, i.e., small power train lag (car responds fast to acceleration commands) and increased engine performance. Acceleration commands, $u(t)$, and acceleration measurements, $y(t)$, are periodically sampled and sent to the cloud in real-time. We require the cloud to compute our average acceleration on the highway, but we do not wish it to classify our car as a sports car. To this end, we use the model of a second vehicle of the form \eqref{exp_1}, with $\tau = 0.60$ and $\beta = 0.70$, and map acceleration commands/meaurements to its behavior. With this, we want the cloud to classify our vehicle as having an inefficient engine (aerodynamic design) and a large power train lag, i.e., an affordable average car. We exactly discretize both systems at the sampling time-instants with sampling interval of $0.1$ seconds. The resulting discrete-time systems are of the form \eqref{1a} with $(A_1,B_1,C_1)$ and $(A_2,B_2,C_2)$ given by \vspace{1mm}
\begin{align*}
\left\{\begin{array}{ll}
\begingroup \renewcommand*{\arraycolsep}{3pt}
\begin{pmatrix}[c|c|c]
  A_1 & B_1 & C_1^\top
\end{pmatrix} =
\begin{pmatrix}[ccc|c|c]
  1 & 0.1 & 0.0009000& 0.0061499 & 0\\
  0 & 1.0 & 0.0099995& 0.1350010 & 0\\
  0 & 0.0 & 0.0000453& 1.4999300 & 1
\end{pmatrix},\endgroup\\[6mm]
\begingroup \renewcommand*{\arraycolsep}{3pt}
\begin{pmatrix}[c|c|c]
  A_2 & B_2 & C_2^\top
\end{pmatrix} =
\begin{pmatrix}[ccc|c|c]
  1 & 0.1 & 0.0047334& 0.0001866 & 0\\
  0 & 1.0 & 0.0921110& 0.0055223 & 0\\
  0 & 0.0 & 0.8464820& 0.1074630 & 1
\end{pmatrix},\endgroup
\end{array} \right.
\end{align*}
and $\mathcal{N} = \{1,2\}$. The target mode is $\rho^\prime = 2$, and the operation mode of the vehicle is $\rho=1$. The driving time is one hour and because the sampling time is 0.01 seconds, the length of the trajectory is $K=36000$. Next, we use the procedure in Section \ref{synth} to synthesize the distorted trajectory to be shared with the cloud. We first seek a solution to the matrix equations in \eqref{reg}. It can be verified that, for the matrices $(A_i,B_i,C_i)$, $i=1,2$, introduced above, the following $\Pi$ and $\Gamma$, and $\Theta = 13.95$, are a solution to \eqref{reg} (the regulator equations):
\vspace{1mm}
\begin{align*}
\left\{\begin{array}{ll}
\begingroup \renewcommand*{\arraycolsep}{3pt}
\begin{pmatrix}[c|c]
  \Pi & \Gamma^\top
\end{pmatrix} =
\begin{pmatrix}[ccc|c]
  1 & -0.038 &  \hspace{3mm}0.001 & \hspace{3mm}0.000\\
  0 &  \hspace{3mm}1.000 & -0.038 & \hspace{3mm}0.000\\
  0 &  \hspace{3mm}0.000 &  \hspace{3mm}1.000 & -7.876
\end{pmatrix}.\endgroup
\end{array} \right.
\end{align*}
We randomly fix the initial condition of system $\Sigma_1$, $x_1(1)$, select $R$ as $R = (-468.99, -130.18,  -13.40)$ (which leads to $\text{eig}[A_2 + B_2R] = \{0.1,0.2,0.3\}$), and extract $(L,S)$ of controller $\bar{u}^1(k)$ in \eqref{controller1} and the initial condition $\bar{x}^1(1)$ of \eqref{target1} using Corollary 1. We close \eqref{target1} with these $\bar{x}^1(1)$ and $\bar{u}^1(k)$. From Theorem 1 and Corollary 1, $\bar{y}^1(k) = y(k) = a(k)$ for all $k \in \mathcal{K}$, and, by construction, the input-output trajectory, $(\bar{u}^1(1),\ldots,\bar{u}^1(K-1),\bar{y}^1(1),\ldots,\bar{y}^1(K))$, belongs to $\mathcal{B}_2$. In Figure \ref{fig1}, we show traces of $y(k)$ and $\bar{y}^1(k)$ (they are indistinguishable), and in Figure \ref{fig2} input trajectories, $u(k)$ and $\bar{u}^1(k)$, are depicted. Next, we use Theorem 2 and Corollary 2 to compute $\bar{x}^1(1)$ and  $\bar{u}^2(k)$. We aim at keeping the average acceleration of the trajectory invariant. Then, the utility matrix $F$ and vector $\mu$ in \eqref{map2} are $F = (1/K)\mathbf{1}_{1 \times K}$ and $\mu = \mathbf{0}$. It can be verified that matrix $(\mathcal{O}_K \hspace{1mm} \mathcal{T}_K \hspace{1mm} (F^+F - I_{Km}))$ in Theorem 2 is full row rank. Then, there exist an infinite number of solutions to \eqref{map3}, which implies that we can induce arbitrarily large distortion to trajectories without affecting their utility (see Remark 2). We randomly choose a solution to \eqref{map3}, and compute $\bar{x}^2(1)$ of \eqref{target2} and the sequence of controllers $\{\bar{u}^2(1),\ldots,\bar{u}^2(K-1)\}$ using Corollary \ref{corollary2}. Finally, $\bar{u}(k) = \bar{u}^1(k) + \bar{u}^2(k)$ and $\bar{y}(k) = \bar{y}^1(k) + \bar{y}^2(k)$ are computed, and $(\bar{u}(k),\bar{y}(k))$ is sent to the cloud in real-time (see Section \ref{synth} for details). Both $Y^K$ and $\bar{Y}^K$ lead to the same utility $FY^K = F\bar{Y}^K = 0.0234$. In Figure \ref{fig3}, we show traces of $y(k)$ and $\bar{y}(k)$, and in Figure \ref{fig4} input trajectories, $u(k)$ and $\bar{u}(k)$, are depicted.

\section{Conclusion}

We have proposed a new formulation for dealing with privacy problems in cyber-physical systems. In particular, for a class of CPSs, we have addressed the problem of performing computations over the cloud without revealing private information about the structure and operation of the system. A distorting mechanism (based on output regulation techniques) that ensure CPSs data privacy and utility invariance has been proposed. We have provided simulation results to test the performance of our tools.


\bibliographystyle{IEEEtran}
\bibliography{ifacconf32}

\end{document}